\documentstyle[12pt,epsf]{article}
\newcommand{\bm}[1]{\mbox{\boldmath #1}}

\newcommand{\SR}{S_{\hbox{{\scriptsize Regge}}}}
\newcommand{\SLCS}{S_{\hbox{{\scriptsize LCS}}}}

\newcommand{\ZPR}{Z_{\hbox{{\scriptsize PR}}}}
\newcommand{\plaq}{\displaystyle \prod _{\partial \tilde{P}(l)} \hspace*{-6.4mm} \bigcirc \ U}
\newcommand{\sixj}[6]{
\renewcommand{\arraystretch}{1}
  \left\{
   \begin{array}{ccc}
    #1 & #2 & #3 \\
    #4 & #5 & #6
   \end{array}
\right\}}
\newcommand{\threej}[6]{
\renewcommand{\arraystretch}{1}
 \left(
  \begin{array}{ccc}
    #1 & #2 & #3 \\
    #4 & #5 & #6
  \end{array}
 \right)}
\renewcommand{\theequation}{\arabic{section}.\arabic{equation}}

\makeatletter
\def\eqnarray{%
 \stepcounter{equation}%
 \let\@currentlabel=\theequation
 \global\@eqnswtrue
 \global\@eqcnt\z@
 \tabskip\@centering
 \let\\=\@eqncr
 $$\halign to \displaywidth\bgroup\@eqnsel\hskip\@centering
 $\displaystyle\tabskip\z@{##}$&\global\@eqcnt\@ne
 \hfil$\displaystyle{{}##{}}$\hfil
 &\global\@eqcnt\tw@$\displaystyle\tabskip\z@{##}$\hfil
 \tabskip\@centering&\llap{##}\tabskip\z@\cr}
\makeatother
\begin{document}

\renewcommand{\theequation}{\arabic {section}.\arabic{equation}}
\setlength{\baselineskip}{7mm}
\begin{titlepage}
\begin{flushright}
EPHOU-99-002 \\
February, 1999
\end{flushright}
 
\vspace{15mm}
\begin{center} 
{\Large Lattice Chern-Simons Gravity   \\}
{\Large via Ponzano-Regge Model }\\
\vspace{1cm}
{\bf {\sc Noboru Kawamoto$^{1)}$
 \footnote{kawamoto@particle.sci.hokudai.ac.jp},
 Holger Bech Nielsen$^{2)}$
 \footnote{HBECH@nbivms.nbi.dk}
 and Noriaki Sato$^{1)}$
 \footnote{noriaki@particle.sci.hokudai.ac.jp}}}\\
{\it{ $^{1)}$Department of Physics, Hokkaido University }}\\
{\it{ Sapporo, 060-0810, Japan}}\\
{\it{ $^{2)}$Niels Bohr Institute }}\\
{\it{ Blegdamsvj 17, Copenhagen, Denmark}}\\
\end{center}
\vspace{2cm}

\begin{abstract}
We propose a lattice version of Chern-Simons gravity and show that 
the partition function coincides with Ponzano-Regge model 
and the action leads to the Chern-Simons gravity in the continuum limit.
The action is explicitly constructed by lattice dreibein and spin connection 
and is shown to be invariant under lattice local Lorentz transformation 
and gauge diffeomorphism.
The action includes the constraint which can be interpreted as a gauge fixing 
condition of the lattice gauge diffeomorphism. 
\end{abstract}

\vspace{2 cm}

\end{titlepage}

\setcounter{equation}{0}
\section{Introduction}

It is not obvious that the string is the only formulation leading to 
the quantum gravity. 
In fact, two dimensional quantum gravity was formulated by a lattice 
gravity, the dynamical triangulation of random surface. 
On the other hand, three dimensional Einstein gravity was successfully 
formulated by the Chern-Simons action even at the 
quantum level\cite{Witten}.
It is natural but nontrivial expectation that the three dimensional 
gravity will be formulated on the lattice. 
In particular it is natural to ask 
how one can formulate the Chern-Simons gravity on the lattice.

Ponzano and Regge (P-R) proposed
a 3-dimensional lattice gravity model
based on 6-$j$ symbols\cite{Ponzano-Regge} about 30 years ago.
At the early '90s Turaev and Viro (T-V) proposed 
$q$-deformed version of the P-R model\cite{Turaev-Viro}
and then Turaev pointed out that the partition function of the 
T-V model is the square of the partition function of $SU(2)$ 
Chern-Simons gauge theory which is equivalent to the Palatini 
action with a cosmological term\cite{Turaev}\cite{Ooguri-Sasakura}.
On the other hand Ooguri and Sasakura showed that 
the P-R model is equivalent to
the $ISO(3)$ Chern-Simons theory\cite{Ooguri-Sasakura}.
They showed the equivalence by using the wave function of 
Wheeler-DeWitt equation and the knowledge of conformal field 
theory. 
The proof is, however, indirect.

There is also another indirect approach to show the 
equivalence of the P-R model and Chern-Simons gravity.
Vanishing curvature condition is the equation of motion of 
the Chern-Simons gravity and can be used to derive 6-$j$ symbol 
via orthogonality of character\cite{Boulatov}\cite{Fukuma}\cite{ACSM}. 
Thus the equivalence is on the classical level.

Apart from these development two of the present authors (Kawamoto 
and Nielsen) proposed a gravity version of Wilson's lattice gauge 
theory\cite{Kawamoto-Nielsen} where the plaquette action plays a
fundamental role. 
There was an independent proposal\cite{Dada} similar to ours.  

In this paper we extend the formulation previously proposed by the 
authors and explicitly construct a lattice Chern-Simons gravity 
by identifying the location of dreibein and spin connection 
on a simplicial lattice manifold. 
After the integration of the dreibein and spin connection, 
we obtain the Ponzano-Regge model. 
We clarify the lattice version of local Lorentz 
transformation and the gauge diffeomorphism.
We then give arguments that the lattice action leads to 
the Chern-Simons action in the continuum limit.

The standard Chern-Simons action is formulated by one form gauge field and 
zero form gauge parameter. 
Since the three dimensional Chern-Simons gravity is formulated only by 
forms, the general coordinate diffeomorphism invariance of the action is 
trivial and should be reflected on the lattice.
The standard Chern-Simons action has been generalized into arbitrary 
dimensions\cite{Kawamoto-Watabiki1} by introducing all the possible 
form degrees. 
It has been analysed that the two and four dimensional generalized 
Chern-Simons actions lead to a two dimensional topological gravity and 
four dimensional topological conformal gravity,respectively, at the 
classical level\cite{Kawamoto-Watabiki2}.
One of the important aim of the current analysis of the lattice Chern-Simons 
gravity is to extend the formulation given here into other dimensions 
including four by using the generalized Chern-Simons actions.

\setcounter{equation}{0}
\section{Brief Summary of Chern-Simons Gravity and Ponzano-Regge Model} 
\subsection{Chern-Simons Gravity}
We first summarize the Chern-Simons gravity formulated by 
Witten\cite{Witten}.
We choose the gauge group as Euclidean version of three dimensional 
Poincare group $ISO(3)$.
Then we define one form gauge field and zero form gauge parameter as 
\renewcommand{\arraystretch}{1.5}
\begin{equation}
\begin{array}{rcl}
  A_\mu&=&e^a_\mu P_a + \omega^a_\mu J_a, \\
 v  &=& \rho^a P_a + \tau^a J_a,
\end{array}
\end{equation}
where $e^a_\mu$ and $\omega^a_\mu$ are dreibein and spin connection, 
respectively, and $\rho$ and $\tau$ are the corresponding gauge parameters. 
The momentum generator $P_a$ and the angular momentum generator $J_a$ 
of $ISO(3)$ satisfy 
\begin{equation}
 [J_a,J_b] = \epsilon_{abc} J^c, ~~~
 [J_a,P_b] = \epsilon_{abc} P^c, ~~~
 [P_a,P_b] = 0.
\end{equation} 
Using the invariant quadratic form which is particular 
in three dimensions, we can define the inner product 
\begin{equation}
 \langle J_a,P_b \rangle = \delta_{ab},
 ~~\langle J_a,J_b \rangle = \langle P_a,P_b \rangle = 0.
\end{equation} 
We then obtain Einstein-Hilbert action of three 
dimensional gravity from Chern-Simons action   
\begin{equation}
 \int \Bigl\langle AdA + \frac{2}{3}A^3 \Bigr\rangle = 
 \int \epsilon^{\mu\nu\rho}e_{\mu a}F^a_{\nu\rho}~d^3x,
\end{equation}  
where 
\begin{equation}
 F^a_{\mu\nu}= \partial_\mu\omega^a_\nu - 
               \partial_\nu\omega^a_\mu +
               \epsilon^a_{~bc}\omega^b_\mu\omega^c_\nu.\label{CSCurvature}
\end{equation} 

The component wise gauge transformation of 
$\delta A_\mu = - D_\mu v$ is given by 
\begin{equation}
\begin{array}{rcl}
 \delta e^a_\mu &=& -D_\mu\rho^a -\epsilon^{abc}e_{\mu b}\tau_c,\\
 \delta \omega^a_\mu &=& -D_\mu\tau^a.\label{CS gauge transformation}
\end{array}
\end{equation} 
At this stage it is important to recognize that 
the local Lorentz transformation is generated by the 
gauge parameter $\tau$
\begin{equation}
\begin{array}{rcl}
 \delta e^a_\mu &=&  -\epsilon^{abc}e_{\mu b}\tau_c,\\
 \delta \omega^a_\mu &=& -D_\mu\tau^a, \label{LLGT}
\end{array}
\end{equation}
while the gauge transformation of diffeomorphism is generated 
by the gauge parameter $\rho$
\begin{equation}
\begin{array}{rcl}
 \delta e^a_\mu &=& -D_\mu\rho^a, \\
 \delta \omega^a_\mu &=& 0.\label{DGT}
\end{array}
\end{equation}

Three dimensional Einstein gravity is thus elegantly 
formulated by Chern-Simons action. 
This is essentially related to the fact that 
the three dimensional Einstein gravity does not include
dynamical graviton and thus can be formulated by the 
topological Chern-Simons action.
The equivalence of the above action and Einstein-
Hilbert action is, however, valid only if the dreibein $e^a_\mu$ 
is invertible.
The quantization and perturbative renormalizability 
around the nonphysical classical background 
$e^a_\mu = 0$ 
is the natural consequence of the formulation. 

It has been pointed out that $ISO(3)$ Chern-Simons gravity action 
is equivalent to the Palatini action for three dimensional gravity, 
which is essentially the three dimensional version of 
$BF$ theory\cite{Ashtekar}. 
In $BF$ theory $B$ is a one form on the three dimensional manifold 
taking values in ${\cal L}^*_G$ and $F$ is a curvature two form on 
the manifold taking values in ${\cal L}_G$, where ${\cal L}^*_G$ 
is the dual algebra of ${\cal L}_G$.
In the present case ${\cal L}^*_G \oplus {\cal L}_G$ coincides with 
the Lie algebra of $ISO(3)$ group. 
Due to the algebraic dual nature of $B$ and $F$, the gauge transformations 
of the Palatini action include the same gauge transformations (\ref{LLGT}) 
and (\ref{DGT}) as $ISO(3)$ Chern-Simons gauge theory.

\subsection{Ponzano-Regge Model}

Ponzano and Regge noticed that angular momenta of 6-$j$ symbol 
can be identified as link lengths of a tetrahedron.
\begin{figure}
\begin{center}
 \begin{minipage}[b]{0.4\textwidth}
 \epsfxsize=\textwidth \epsfbox{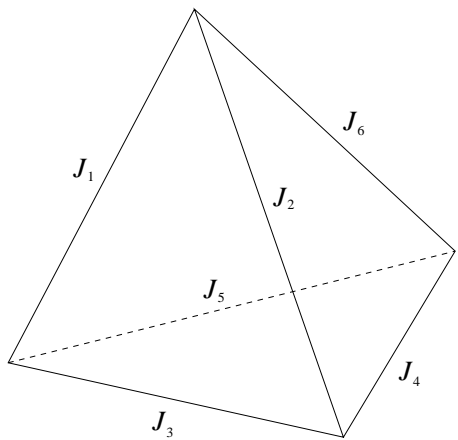} 
 \end{minipage}
\end{center}
 \caption{tetrahedron with angular momenta on the links}
 \label{fig:colored tetra}
\end{figure}
In particular they showed the following approximate relation: 
\begin{equation}
 (-1)^{\sum_{i=1}^{6} J_{i}} 
  \sixj{J_1}{J_2}{J_3}{J_4}{J_5}{J_6}
  \sim 
  \frac{1}{\sqrt{12 \pi V}} \cos \left( \SR + \frac{\pi}{4} \right) 
  \quad (\hbox{all } J_i \gg  1) ,
  \label{P-R formula}
\end{equation}
where $\SR$ is the Regge action of Regge calculus\cite{Regge} 
for a tetrahedron having link length $J_k$ $(k=1\sim 6)$ 
which correspond to the angular momentum of the corresponding 6-$j$ symbol 
and $V$ is the volume of the tetrahedron.
Based on this observation they proposed the following partition function: 
\begin{equation}
 \ZPR = \lim_{\lambda \rightarrow \infty} \sum_{J \leq \lambda} 
  \prod_{\hbox{\tiny vertices}} \Lambda (\lambda) ^{-1} 
  \prod_{\hbox{\tiny edges}} (2J+1) 
  \prod_{\hbox{\tiny tetrahedra}} (-1)^{\sum J_i} 
    \sixj{J_1}{J_2}{J_3}{J_4}{J_5}{J_6}
\label{ZofPR}.
\end{equation}
Thus the partition function $\ZPR$ is 
the product of the partition function of each tetrahedron which 
reproduces the cosine of the Regge action in contrast with 
the exponential of the Regge action in Regge calculus.
There is an argument about the origin of the cosine, that right 
and left handed contributions of the general coordinate frames 
contribute separately and thus the summation of the exponential 
with the different sign factor for the Regge action appears.
It is thus natural to expect that this 
action leads to a gravity action.

\begin{figure}
\begin{center}
 \begin{minipage}[b]{0.4\textwidth}
 \epsfxsize=\textwidth \epsfbox{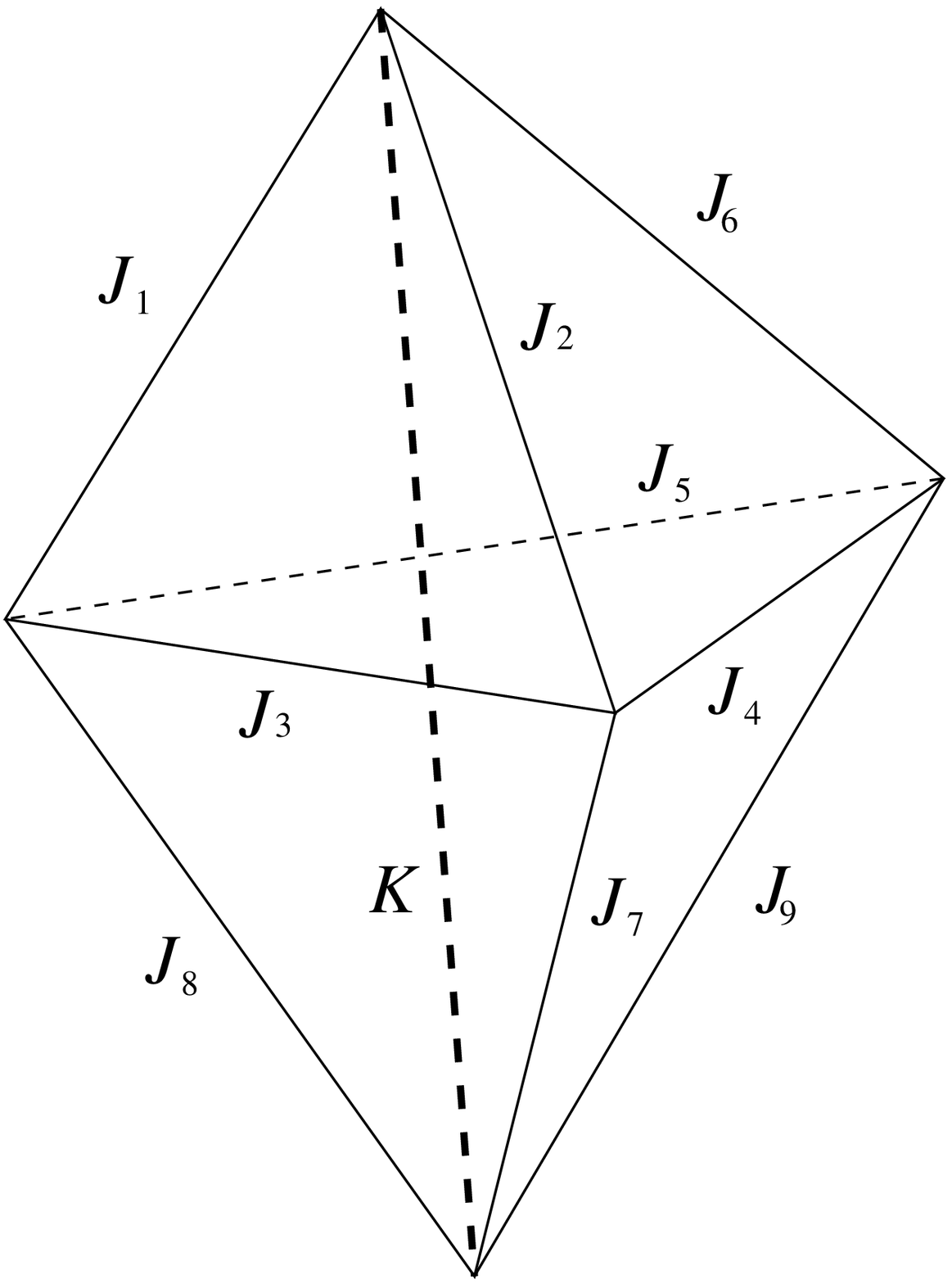} 
 \end{minipage}
 \hspace*{2cm}
  \begin{minipage}[t]{0.4\textwidth}
 \epsfxsize=\textwidth \epsfbox{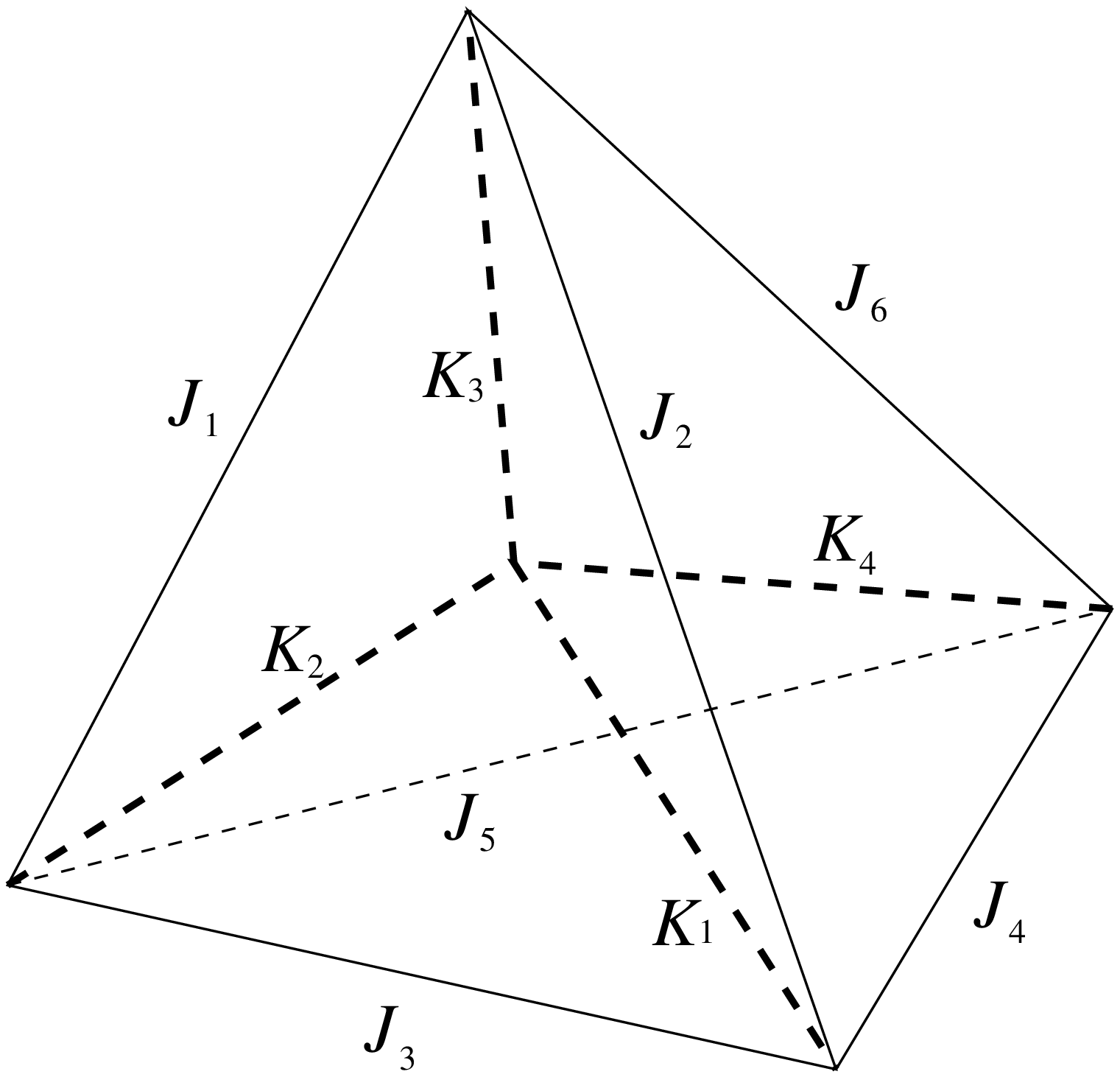} 
 \end{minipage}
\end{center}
 \caption{2-3 move and 1-4 move}
 \label{fig:Alexander move}
\end{figure}

Important characteristic of the Ponzano-Regge action is that 
it has a topological nature on a simplicial manifold. 
The action is invariant under the following 2-3 and 1-4 
Alexander moves. 
The 2-3 and 1-4 moves are related to the following 6-$j$ relations:
\begin{eqnarray}
 && \sum_{K} (-1)^{K+\sum_{i=1}^{9}J_i} (2K+1) 
  \sixj{J_1}{J_8}{K}{J_7}{J_2}{J_3} \sixj{J_7}{J_2}{K}{J_6}{J_9}{J_4}
  \sixj{J_6}{J_9}{K}{J_8}{J_1}{J_5} 
    \nonumber \\
 &=& \sixj{J_3}{J_4}{J_5}{J_6}{J_1}{J_2} \sixj{J_3}{J_4}{J_5}{J_9}{J_8}{J_7},
  \label{2-3move}
\end{eqnarray}
and 
\begin{eqnarray}
&&  \sum_{K_i} \left[ \prod_{i=1}^{4} (2K_i + 1) \right]
 (-1)^{\sum K_i} \Lambda (\lambda)^{-1} 
  \sixj{J_1}{J_2}{J_3}{K_1}{K_2}{K_3}
  \sixj{J_4}{J_6}{J_2}{K_3}{K_1}{K_4} \nonumber \\
&& \hspace{2cm} \times
  \sixj{J_3}{J_4}{J_5}{K_4}{K_2}{K_1}
  \sixj{J_1}{J_5}{J_6}{K_4}{K_3}{K_2} 
= (-1)^{\sum J_i} \sixj{J_1}{J_2}{J_3}{J_4}{J_5}{J_6}.
\label{1-4move}
\end{eqnarray}
The geometrical correspondence of 2-3 and 1-4 moves with 
two tetrahedra into three tetrahedra and one tetrahedron 
into four tetrahedra is obvious from 
Fig.\ref{fig:Alexander move}.
In the formula of 1-4 move there appears the following infinite sum 
which is then introduced as a regularization factor in the denominator 
with a cutoff $\lambda$:  
\begin{eqnarray}
 \Lambda (\lambda)
&=& \frac{1}{2 J_1 + 1}  
\!\!\!\!\!\!\!\!\!\!
  \sum_{ 
  \mbox{ \scriptsize $
  \begin{array}{c} 
   K_2,K_3 \leq \lambda,\\ 
   |K_2 - K_3| \leq J_1 \leq K_2 + K_3
  \end{array} $}
   }
\!\!\!\!\!\!\!\!\!\!
(2 K_2 + 1)(2 K_3 + 1) \nonumber  \\
&=& \sum_{J=0}^{\lambda} (2J+1)^2  
\sim \frac{4 \lambda^3}{3} ~~ (\lambda \rightarrow \infty)
\label{Lambda}.
\end{eqnarray}

It is known that these two Alexander moves reproduce any three dimensional 
simplicial manifold.
Thus the partition function $\ZPR$ is invariant 
under the variation of metric and is expected to be topological.  

In this paper we show that the continuum limit of the lattice 
Ponzano-Regge model leads to the Chern-Simons gravity by explicitly 
constructing lattice gauge gravity model.

\setcounter{equation}{0}
\section{Lattice Chern-Simons Gravity Action }  \label{Action}

We consider a three-dimensional piece-wise linear simplicial manifold 
which is composed of tetrahedra.
In 3-dimensional Regge calculus curvature is concentrated on the links of 
tetrahedra.
We intend to formulate a lattice gravity theory in terms of gauge variables, 
dreibein $e$ and spin connection $\omega$.
In analogy with the lattice gauge theory where link variables surrounding a 
plaquette induce a gauge curvature, it was proposed in \cite{Kawamoto-Nielsen} 
and \cite{Dada} 
that dual link variables $U(\tilde{l}) = e^{\omega(\tilde{l})}$ 
located at the boundary of a dual plaquette $\tilde P$ 
($\tilde{l} \in \partial\tilde{P}(l)$) 
associated to an original link $l$ 
induce the curvature of the gravity theory.
It was further pointed out that the dreibein $e^a(l)$ is located on the 
original link $l$.

We propose to use a lattice version of Chern-Simons action,
which is a modified version of the one in \cite{Kawamoto-Nielsen},
and show that the length of dreibein $e(l)$ is naturally discretized.
In the Chern-Simons formulation,
the dreibein $e^a$ and spin connection $\omega ^{ab}$ are Lie algebra valued 
gauge fields. 
For a moment we consider a Euclidean version of three-dimensional local 
Lorentz group $SO(3)$ and discuss $SU(2)$ case later.

\begin{figure}[t]
\begin{center}
 \begin{minipage}[c]{0.4\textwidth}
 \epsfxsize=\textwidth \epsfbox{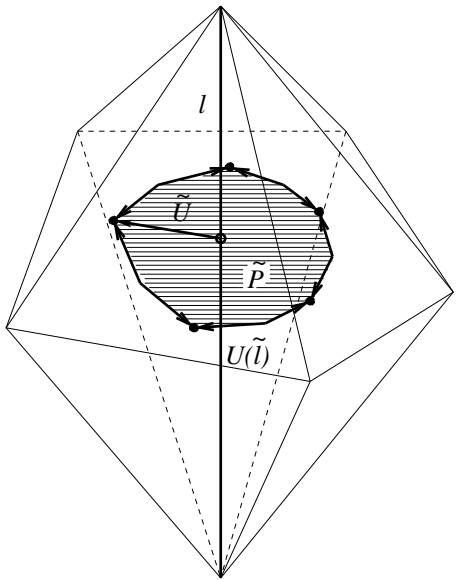} 
 \end{minipage}
\end{center}
 \caption{dual link variables on $\partial \tilde{P}$}
 \label{fig:action}
\end{figure}

Here we slightly modify the formulation given above in order that each 
tetrahedron gets independent contribution to the partition function 
and at the same time the orientability could be naturally accommodated.  
We divide the dual link, which connects the centers of neighbouring 
tetrahedra, into two links by the center of mass of the common triangle 
of the neighbouring tetrahedra.
We may keep to use the terminology of dual plaquette and dual link 
even for those modified plaquettes and links. 
Correspondingly we put different link variables $U$ for the doubled dual links.
We then assign the directions of $U$-links inward for each tetrahedron 
as shown in Fig.\ref{fig:action}. 

Using these variables,
we consider the following lattice version of Chern-Simons action
on the simplicial manifold,
\begin{equation}
 \SLCS = \sum_{l} \epsilon_{abc} e^{a}(l)
  \Bigl[ \ln \plaq \Bigr]^{bc}, \label{LCS1} 
\end{equation}
where,
$\partial \tilde{P}(l)$ is a boundary of the $\tilde{P}(l)$,
which is a (dual) plaquette around the link $l$,
and $\displaystyle \plaq $ denotes the product of $U(\tilde{l})$ 
along $\partial \tilde{P}(l)$.
We define the ``curvature'' $F^{ab}(l)$ of the link $l$
by the following equation,
\begin{equation}
 \Bigl[ \plaq \Bigr]^{ab} \equiv \Bigl[ e^{F(l)} \Bigr]^{ab}.
\end{equation}
The leading term of $F$ with respect to the lattice unit is the 
ordinary curvature
$d\omega + \omega \wedge \omega$
similar to the ordinary lattice gauge theory.

It is the unusual point of this action (\ref{LCS1})
that we have taken logarithm of the Wilson loop $U$ along 
$\partial \tilde{P}(l)$ to extract the curvature.
Because of this logarithm, however, this action has several 
nice features.
Firstly, this action itself is already hermitian 
in contrast with the ordinary lattice gauge theory 
due to the hermiticity of Lie algebra valued curvature.
Secondly, as we show in the following, the length of dreibein $e$ is 
discretized as a natural consequence of the logarithm.

Classically the Chern-Simons action impose a torsion free condition as an 
equation of motion. 
The torsion free nature is lost at the quantum level since we integrate 
out the dreibein and spin connection.
We now introduce the following vanishing holonomy constraint which relates 
the dreibein and 
spin connection even at the quantum level:
\begin{equation}
 \Bigl[ \plaq \Bigr]^{ab} e^b = e^a. \label{constraint}
\end{equation}
The dreibein $e^a$ associated to a original link may be parallel transported around 
the boundary of the dual plaquette $\partial \tilde{P}(l)$ to the original 
location and yet the direction of the dreibein should not be changed.
We may interpret this constraint as a gauge fixing condition 
of gauge diffeomorphism symmetry which we will explain 
later.
Due to the constraint the group $SO(3)$ becomes ``effectively abelian'', 
i.e. the direction of the rotation associated with the curvature is 
parallel to that of $e^a$.
This can be seen as follows: we can reduce the above constraint to
the following one:
\begin{equation}
  F^{ab} e^b = 0, \label{constraint2}
\end{equation}
hence $F^a \equiv \frac{1}{2} \epsilon ^{abc} F^{bc}$ is parallel
to $e^a$: $e^a \propto F^a$.

Here we should reconsider the constraint (\ref{constraint}). 
Firstly it should be noted that the $\displaystyle\Bigl[ \plaq \Bigr]^{ab}$ 
is an 
element of $SO(3)$ and thus the eigenvalue equation of this element always has 
eigenvalue +1. 
Thus the number of the independent constraints in eq.(\ref{constraint})   
is not three but two.
Taking into account the parallel and anti-parallel nature of $e^a$ and $F^a$ 
in the constraint, we can rewrite the correct constraint equation  
\begin{equation}
\frac{e^3}{|e|} 
\left[ \prod_{a=1}^{2}
 \delta \left( \frac{F^a}{|F|} + \frac{e^a}{|e|} \right)
+ \prod_{a=1}^{2}
\delta \left( \frac{F^a}{|F|} - \frac{e^a}{|e|} \right) \right], 
\label{constraint3}
\end{equation}
where $|e|$ and $|F|$ are length of $e^a$ and $F^a$, respectively.
The coefficient factor $\frac{e^3}{|e|}$ is necessary to keep the 
rotational invariance of the constraint relation, which 
can be easily checked by polar 
coordinate expression of the constraint relation.

Now we show that discreteness of the length of the dreibein $|e|$ 
comes out as a natural consequence of the specific choice of the lattice 
gauge gravity action.
We first introduce the following normalized matrix $I$,
\begin{equation}
 I \equiv I^{a} J_{a} , 
\quad  I^{a} \equiv \frac{F^{a}}{\sqrt{F^{a}F_{a}}},
\end{equation}
here $[J_{a}]_{bc} = i \epsilon_{abc}$ is the generator of $SO(3)$.
This matrix satisfies the following relation,
\begin{equation}
 e^{i \theta I} = 1 - I^2 (1-\cos \theta) + i I \sin \theta,
\end{equation}
then
\begin{equation}
 e^{i 2 \pi n I} = 1, \quad n \in \bm{Z} \label{periodicity1}.
\end{equation}
Using the above relation 
and $F^a \propto e^a$ by the constraint (\ref{constraint}),
we find that our lattice Chern-Simons action $\SLCS$ has 
the following ambiguity:
\begin{eqnarray*}
 \SLCS
  &=& \sum _{l} \epsilon _{abc} e^{a}(l)  
  \left[ \ln e^{F(l)} \right] ^{bc} \\
  &=& \sum _{l} \epsilon _{abc} e^{a}(l)  
  \left[ \ln e^{F(l) + i 2 \pi n I} \right] ^{bc} \\
  &=& \sum _{l} \bigl[ 2 e^{a}(l) F_a (l) + 4 \pi n |e(l)| \bigr] \\
  &=& \SLCS + \sum_{l} 4 \pi n |e(l)|,
\label{ambiguity}
\end{eqnarray*}
here $|e|$ is the length of $e^a$, $|e| \equiv \sqrt{e_a e^a}$.
This ambiguity leads to an ambiguity in the partition 
function
\begin{equation}
 Z = \int {\cal D}U {\cal D}e ~ e^{i S_{\hbox{\tiny LCS}}}
   = \int {\cal D}U {\cal D}e ~
   e^{ i S_{\hbox{\tiny LCS}} + i \sum_{l} 4 \pi n |e|}.
\end{equation}
Imposing the single valuedness of 
$e^{i S_{\hbox{\tiny LCS}}}$, we obtain the constraint 
that $\sum_{l} 2 |e(l)| $ should be integer, or equivalently $|e(l)|$ should 
be half integer.

In the above arguments we have restricted the dual link variables to $SO(3)$. 
If we extend the arguments to $SU(2)$ the discrete nature of the dreibein is 
modified as follows. 
First of all we need to use the triplet representation of $SU(2)$ for the 
dual link variables since the suffix of the color variable should vary from 
1 to 3 to be compatible with our lattice Chern-Simons action (\ref{LCS1}).
We may then use the same generators of $SO(3)$, 
$[J_{a}]_{bc} = i \epsilon_{abc}$, for the triplet representation of $SU(2)$.
This representation is, however, not faithful (injective). 
In other words an element of the triplet representation used by those 
generators and the corresponding element of $SU(2)$ is not one to one but 
one to two correspondent.    
Due to this degeneracy of the representation the periodicity relation 
(\ref{periodicity1}) for $SU(2)$ should be modified to
\begin{equation}
 e^{i 4 \pi n I} = 1, \quad n \in \bm{Z} \label{periodicity2}.
\end{equation}
Accordingly we need to modify factor 2 in the corresponding relations 
in the above, i.e., $4|e(l)|$ should be substituted for $2|e(l)|$ 
in $SU(2)$ case. 

\setcounter{equation}{0}
\section{Gauge Invariance on the Lattice} \label{Gauge Invariance}

The gauge transformations of the continuum Chern-Simons gravity have 
been given by (\ref{CS gauge transformation}) which includes 
the local Lorentz gauge transformation (\ref{LLGT}) and 
the gauge transformation of diffeomorphism (\ref{DGT}).
We first note that  
the dreibein and the curvature defined in (\ref{CSCurvature}) 
transform adjointly under the local Lorentz gauge transformation  
\begin{equation}
\begin{array}{rcl}
 \delta e^a_\mu &=  -\epsilon^{abc}e_{\mu b}\tau_c,\\
 \delta F^a_{\mu\nu} &= -\epsilon^{abc}F^b_{\mu\nu}\tau_c. \label{LLGT2}
\end{array}
\end{equation} 

We consider the lattice version of the local Lorentz gauge parameters
are sitting on the dual sites and the middle of the original links, 
the same point of the dreibein. 
For simplicity we consider here in this section that the 
dual link is not divided into two dual links by the center 
of original triangle.
Then the dual link variable $U(\tilde{l}) = e^{\omega(\tilde{l})}$ 
transforms under the lattice local Lorentz transformation as 
\begin{equation}
U(\tilde{l}) \rightarrow V^{-1} U(\tilde{l}) V', \label{LLLspin}
\end{equation}
where the gauge parameters $V$ and $V'$ are elements of $SO(3)$ 
and located at the end points of dual link $\tilde{l}$.
Defining the matrix form of the dreibein by 
$E^{cb}_\mu(l)=\epsilon^{abc}e_{a\mu}(l)$, 
we can rewrite the lattice Chern-Simons action (\ref{LCS1}) by 
\begin{equation}
 \SLCS = \sum_{l} \hbox{Tr} ( E(l)F(l)), \label{LCSA} 
\end{equation}
where $\displaystyle F(l)^{ab}=\Bigl[ \ln \plaq \Bigr]^{ab}$. 

Corresponding to the continuum local Lorentz transformation, we 
can define the lattice version of local Lorentz transformation of
$E(l)$ and $F(l)$ according to (\ref{LLLspin}) 
\begin{equation}
 \begin{array}{ccc}
     E(l) & \rightarrow & V^{-1}E(l) V,  \\  
     F(l) & \rightarrow & V^{-1}F(l) V.
\end{array}
\label{LLLGT}
\end{equation}
It is obvious that the lattice Chern-Simons action (\ref{LCSA}) 
is invariant under the lattice local Lorentz transformation.

There are, however, some subtleties on the gauge invariance of the
lattice action.
In defining the lattice curvature  
$\displaystyle F(l)^{ab}=\Bigl[ \ln \plaq \Bigr]^{ab}$, we need to choose a 
starting and ending dual site of the product $\displaystyle \plaq$ to define 
the lattice curvature. 
See Fig.\ref{fig:action}.
We then need to bridge between the dual site and the center of 
the original link of $e(l)$ by new link variables $\tilde{U}$ and 
$\tilde{U}^{-1}$.
The action associated with this particular dual plaquette is 
\begin{equation}
 \begin{array}{rcl}
\displaystyle
\hbox{Tr} \Bigl( E(l)\tilde{U}\Bigl[\ln \plaq \Bigr] \tilde{U}^{-1} \Bigr)&=& 
\hbox{Tr} \Bigl( V^{-1} E(l)V V^{-1} \tilde{U}V'V'^{-1} 
\Bigl[ \ln \plaq \Bigr] V'V'^{-1}\tilde{U}^{-1}V \Bigr) \\
&\rightarrow& \hbox{Tr} \Bigl( E'(l) \Bigl[ \ln \plaq' \Bigr] \Bigr).
\end{array}
\label{Utilde}
\end{equation}
Then the newly introduced link variable transforms similar as the dual link 
variable $\tilde{U} \rightarrow V^{-1} \tilde{U} V'$,
where $V$ and $V'$ are 
located on the the center of 
the original link and the dual site, respectively.  
We can, however, use one of the gauge parameters, say the one at center of 
the original link, $V^{-1}$,  
to tune in such a way that this link variable leads to a trivial factor 
$\tilde{U} \rightarrow V^{-1} \tilde{U} V' \rightarrow 1$. 
We then redefine the matrix form of the dreibein 
$V^{-1}E(l)V = E'(l)$ and
the curvature $V'^{-1} \Bigl[ \ln \plaq \Bigr] V' = \ln \plaq '$.
We can thus gauge away the variable $\tilde{U}$.
This can be allowed since we have enough gauge parameters on the 
dual sites and on the original links because of the geometrical reason 
that our simplicial 
manifold is constructed out of tetrahedra.     
We can now identify the final expression of (\ref{Utilde}) as 
the one in (\ref{LCSA}).
Accepting this arguments we have confirmed that our lattice 
Chern-Simons action is invariant under lattice local Lorentz 
transformation.

\begin{figure}
\begin{center}
 \begin{minipage}[b]{0.4\textwidth}
 \epsfxsize=\textwidth \epsfbox{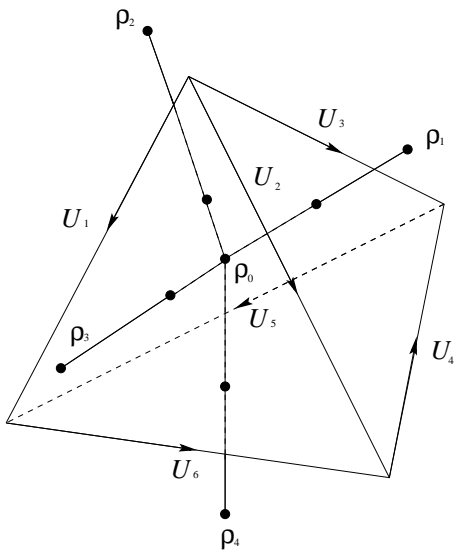} 
 \end{minipage}
\end{center}
 \caption{dual tetrahedron : $\rho _{k}$ $(k=0 \sim 4)$ at original site,
 $U_k$ $(k = 1 \sim 4)$ on dual link}
 \label{fig:Bianchi}
\end{figure}

The continuum Chern-Simons gravity action is invariant under the 
gauge transformation of diffeomorphism (\ref{DGT}) which transforms 
dreibein $e^a_\mu$ but not spin connection $\omega^a_\mu$. 
We try to identify the lattice counter part of this gauge 
transformation and show the gauge invariance. 
The gauge invariance of the continuum action by the gauge diffeomorphism 
(\ref{DGT}) can be shown by using Bianchi identity.
Here we first formulate the lattice version of integrated Bianchi identity 
\begin{equation}
\int_M DF= \int_{\partial M} F + \int_ M[\omega,F] =0 \label{BI}.
\end{equation} 

For a three dimensional simplicial manifold randomly constructed 
from tetrahedra, several original links could be connected to 
an original site. 
Since an original link $l$ is dual to the corresponding 
dual plaquette $\tilde{P}(l)$, an original site is surrounded 
by dual plaquettes which topologically construct $S^2$ sphere.
In general the sphere could take arbitrary shape. 
For simplicity we here assume that the sphere is a tetrahedron.
In this case the original site is in the center of the tetrahedron 
and the triangles of the tetrahedron are the dual plaquettes.
Dual link variables $U_k$ $(k=1\sim 6)$ are located on the dual links, 
the edges of the tetrahedron, where the directions of the dual link 
variables can be arbitrarily chosen. 
Original links $l_k$ $(k=1\sim 4)$ are sticking from the center through 
the triangles. 
See Fig.\ref{fig:Bianchi}.

We first note the following identity:
\begin{equation}  
  \prod U \equiv U_2U_4U^{-1}_3U_3U_5U^{-1}_1U_1U_6U^{-1}_2
                 U_2U^{-1}_6U^{-1}_5U^{-1}_4U^{-1}_2 = 1. \label{ID}
\end{equation}
We now define 

\begin{equation}
 \begin{array}{ccc}
\ln(U_2U_4U^{-1}_3)=F_1,&~~& \ln(U_3U_5U^{-1}_1)=F_2, \\
\ln(U_1U_6U^{-1}_2)=F_3,&~~&
\ln(U_2U^{-1}_6U^{-1}_5U^{-1}_4U^{-1}_2)=F'_4. \label{LCV}
\end{array}
\end{equation}

Due to the Baker-Hausdorff formula the above relations lead 
\begin{equation}
0 = \ln (\prod U) = \sum_{k=1}^3 F_k + F'_4 +
                      \frac{1}{2}\sum_{
		      \mbox{\scriptsize $
		      \renewcommand{\arraystretch}{1}
		      \begin{array}{c}
		       i<j \\ i,j=1,2,3
		      \end{array}$}
		      }[F_i,F_j] +
                      \frac{1}{2}\sum_{k=1}^3[F_k,F'_4] + \cdots.
\end{equation}
On the other hand the lattice version of Bianchi identity (\ref{BI}) can 
be written 
\begin{equation}  
\sum_{k=1}^3 F_k + F'_4 +
                \sum_{k=1}^3[\Omega_k,F_k] + [\Omega_4,F'_4] = 0. \label{LBI}
\end{equation}
Here we have defined irregular lattice curvature $F'_4$ which has 
$U_2$ and $U^{-1}_2$ in the product of dual link variables.
This type of irregular definition is unavoidable in deriving lattice 
Bianchi identity due to the geometrical reason leading to the type of 
identity of (\ref{ID}). 
We can then identify  
\begin{equation}
\renewcommand{\arraystretch}{2}
 \begin{array}{lcl}
\displaystyle
 \Omega_1 = -\frac{1}{4}(F_2+F_3+F'_4)+\cdots, & ~ &
\displaystyle
 \Omega_2 = -\frac{1}{4}(-F_1+F_3+F'_4)+\cdots, \\
\displaystyle
 \Omega_3 = -\frac{1}{4}(-F_1-F_2+F'_4)+\cdots, & ~ &
\displaystyle
 \Omega_4 = -\frac{1}{4}(-F_1-F_2-F_3)+\cdots. \label{Omega}
\end{array}
\end{equation}

In comparing the continuum Bianchi identity (\ref{BI}) with the lattice 
version (\ref{LBI}), we notice that $[\Omega_k,F_k]$ term is the integrand 
in the volume 
integration. 
$\Omega_k$ should thus be defined as an average spin connection inside 
the dual volume, the tetrahedron in the present case. 
Since the curvature itself could be interpreted as an average spin connection 
on a dual plaquette, the $\Omega_k$ defined in (\ref{Omega}) 
is the particular average of the curvature and thus can be 
interpreted as the average spin connection in the dual volume with respect 
to $F_k$.

We now show the lattice gauge diffeomorphism invariance of 
the lattice Chern-Simons action (\ref{LCSA}). 
We first consider the term related to the fourth link $l_4$ and make 
local Lorentz transformation
\begin{equation}
 \begin{array}{rcl}
\hbox{Tr} \Bigl( E(l_4)F(l_4) \Bigr)&=& 
\hbox{Tr} \Bigl( V^{-1} E(l_4)V V^{-1}F(l_4)V \Bigr) \\
&\rightarrow& \hbox{Tr} \Bigl( E'(l_4)U_2 
F(l_4)U_2^{-1} \Bigr)= \hbox{Tr} \Bigl( E'(l_4)F'(l_4) \Bigr),
\end{array}
\end{equation}
where $F(l_4)\equiv F_4=\ln(U^{-1}_6U^{-1}_5U^{-1}_4)$.
Here we take the gauge choice $V=U_2^{-1}$ and further redefine 
$V^{-1} E(l_4)V = E'(l_4)$, 
we obtain the final expression. From now on we identify $E'(l_4)=E(l_4)$.  
In this way we can introduce the unusual definition of the curvature 
$F'_4$ of (\ref{LCV}) in the lattice Chern-Simons action. 
Hereafter we rename $F'_4$ as $F_4$.  

The lattice version of the gauge transformation of diffeomorphism (\ref{DGT})
can be given by using the $\Omega_k$ defined above
\begin{equation}
\delta E_k(l_k) = -\rho_k + \rho_0 - [\Omega_k,\rho_0]  \label{LDGT},
\end{equation} 
where $\rho_k$ is the matrix gauge parameter,
$(\rho_k)^{ab} \equiv \epsilon^{bac} \rho^c_k $.

Then the lattice gauge transformation of diffeomorphism for the 
lattice action (\ref{LCSA}) leads 
\begin{eqnarray}
 \delta \SLCS &=& \sum_{l} \hbox{Tr} \Bigl( \delta E(l)F(l) \Bigr) \nonumber \\
                &=& \sum_{k} \hbox{Tr} \Bigl\{ (-\rho_k + \rho_0 - 
                    [\Omega_k,\rho_0]) F_k \Bigr\} + \cdots  \nonumber \\
                &=& \sum_{k} \hbox{Tr} \Bigl\{ \rho_0
                   \Bigl(\sum_{k=1}^4 F_k
		   + \sum_{k=1}^4 [\Omega_k,F_k]\Bigr) \Bigr\}
                   + \cdots \nonumber \\
                &=& 0, \label{LDGTA}       
\end{eqnarray}
due to the Bianchi identity (\ref{LBI}).
Here we have used the following relation:
\begin{equation}
 \hbox{Tr}\Bigl( [\Omega_k,\rho_0] F_k \Bigr) =
 - \hbox{Tr} \Bigl( \rho_0[\Omega_k,F_k] \Bigr).
\end{equation}
We have thus completed the proof of the invariance of the lattice 
Chern-Simons action under the lattice gauge transformation of diffeomorphism.

We now point out that the constraint (\ref{constraint}) or equivalently 
(\ref{constraint2}) breaks the lattice gauge diffeomorphism while 
the lattice Chern-Simons action itself is invariant, as is shown above.  
The lattice dreibein is transformed but the lattice curvature is not 
transformed under the lattice gauge transformation of the diffeomorphism. 
The precise expression of the constraint (\ref{constraint3}) tells us that 
the dreibein $e^a$ can be rotated by using two gauge parameters of the 
gauge transformation of diffeomorphism to be parallel or anti-parallel to 
the curvature $F^a$. 
The length of the dreibein is discretized and thus the third gauge parameter 
can be exhausted. 
In this sense we can identify the equivalent constraint,
(\ref{constraint}), (\ref{constraint2}), (\ref{constraint3}) as
a gauge fixing condition of 
the lattice gauge transformation of diffeomorphism. 

\setcounter{equation}{0}
\section{Calculation of Partition Function} \label{Integration}

In the previous section we have found that the length of dreibein 
is discretized to half integer for $SO(3)$ and a quarter for $SU(2)$.
To be specific we restrict our arguments for $SO(3)$ for a moment. 
In order to accommodate the discrete nature of the length of the dreibein, 
we first note an identity 
\begin{equation}
  \int^{|e_f|}_{|e_i|} d|e| ~=~ 
  \int^{|e_f|}_{|e_i|}
  \frac{1}{2}\sum_{J=0}^{\infty} \delta \left( |e| - \frac{J}{2} \right) d|e|, 
\end{equation}
where $|e_f|$ and $|e_i|$ are half integer.
We can thus safely insert the delta function constraints without changing the 
value of the partition function.

Then the total partition function is 
\begin{eqnarray}
 Z &=& \int {\cal D}U  \prod_{l} Z_{l}, \\ 
 Z_{l} &=& \int d^3 e ~\frac{e^3}{|e|}
  \left[ \prod_{a=1}^{2}
   \delta \left( \frac{F^a}{|F|} + \frac{e^a}{|e|} \right)
  + \prod_{a=1}^{2}
  \delta \left( \frac{F^a}{|F|} - \frac{e^a}{|e|} \right)
  \right] \nonumber \\ 
 && \hspace*{1cm} \times
  \frac{1}{2}\sum_{J=0}^{\infty} \delta \left( |e| - \frac{J}{2} \right)
  e^{2 i e^a F^a},
\end{eqnarray}
where $Z_{l}$ is the partition function
associated with a link $l$.

\subsection{$e$ integration}

Due to the rotational invariance of the constraints, we can take 
$e^3$ as the third direction of local Lorentz frame without loss of 
generality.
We can then evaluate $e^a$ integral of $Z_{l}$ immediately
thanks to the delta functions
\begin{eqnarray*}
 Z_{l}
  &=& \int d^3 e ~ |e|^2 \frac{e^3}{|e|}
  \left[ \prod_{a=1}^{2}
   \delta \left( e^a + |e| \frac{F^a}{|F|} \right)
  + \prod_{a=1}^{2}
  \delta \left( e^a - |e| \frac{F^a}{|F|}\right)
  \right] 
  \frac{1}{2}\sum_{J} \delta \left( |e| - \frac{J}{2} \right)
  e^{2 i e^a F^a} \\
  &=& \frac{1}{2}\sum_{J} \left( \frac{J}{2} \right)^2
  \left( e^{2i\frac{J}{2} |F|} + e^{-2i\frac{J}{2} |F|} \right) \\
  &=& \sum_{J} \frac{1}{4} J^2 \cos(J|F|).
\end{eqnarray*}
Using the following formula for the character $\chi_J$
of the spin-$J$ representation of $SO(3)$,
\begin{equation}
 \chi _{J} (e^{i \theta^a J_a})
  = \chi _{J} (|\theta|)
  = \frac{\sin\left( (2J+1) \frac{|\theta|}{2} \right)}
  {\sin \left(\frac{|\theta|}{2} \right)},
\end{equation}
where $|\theta|$ is the length of $\theta^a$,
we find 
\begin{eqnarray}
 \chi _{J}(|F|) - \chi _{J-1}(|F|)
 &=& 2 \cos(J|F|). 
\end{eqnarray}
Hence we can naively calculate the link partition function,
\begin{eqnarray*}
 Z_{l}
  &=& \sum_{J=1}^{\infty} \frac{1}{8} J^2 (\chi _{J} - \chi _{J-1}) \\
  &=& \frac{1}{8}
      \left[ \sum_{J=0}^{\infty} J^2 \chi _{J}
           - \sum_{J=0}^{\infty} (J+1)^2 \chi _{J}
      \right] \\
  &=& \frac{1}{8}
      \sum_{J=0}^{\infty} \left[J^2 - (J+1)^2 \right]\chi _{J} \\
  &=& -\frac{1}{8}
      \sum_{J=0}^{\infty} (2J+1) \chi _{J}.
\end{eqnarray*}
This calculation is not precise, because 
the summation is not convergent.
We need to show that there is a regularization procedure which leads to 
a validity of the above calculation after the regularization.

We propose to use the heat kernel regularization. 
We first consider the following heat equation:
\begin{equation}
  \bigtriangleup  K(U,U'; t)
= \frac{\partial}{\partial t} K(U,U'; t),
 \quad
 \lim_{t \rightarrow 0} K(U,U'; t) = \delta (U,U'), \label{heat equation}
\end{equation}
where the laplacian is defined on the group manifold,
$K(U,U'; t)$ is the heat kernel
and $\delta (U,U')$ is the delta function.
Then the regularized character will be given by 
\begin{equation}
 \chi _J(U;t) = \int dU' \chi _J(U') K(U,U';t). \label{regularization of chi}
\end{equation}

All these quantities are defined on the group manifold $SO(3)$.
In particular the laplacian on the group manifold 
will be related in general to the 2nd Casimir operator, the square of the 
angular momentum in case of $SO(3)$. 
Hence the character of spin-$J$ representation $\chi_J$, which is essentially 
the trace of the matrix representation, is the eigenfunction of 
the laplacian with the eigen value $-J(J+1)$,
\begin{equation}
       \bigtriangleup \chi_J = -J(J+1) \chi_J.
        \label{eigen equation}
\end{equation}
Noting the completeness of the character 
\begin{equation}
      \delta (U,U') = \sum_{J} \chi_J(U) \chi_J(U'),
       \label{delta function}
\end{equation}
we can immediately obtain the heat kernel solution
\begin{equation}
 K(U,U'; t) = \sum_{J}  e^{-J(J+1)t}  \chi_J(U) \chi_J(U').
\end{equation}
Substituting the heat kernel solution into (\ref{regularization of chi}) 
and using the orthogonality of the character, 
\begin{equation}
 \int dU \chi_I(U) \chi_J(U) = \delta_{IJ},
\end{equation}
we obtain an explicit form of regularized character
\begin{equation}
 \chi _{J}(U;t) = e^{-J(J+1)t} \chi _J(U).
\end{equation}
The summation is now convergent and should be replaced by 
\begin{eqnarray*}
 Z_{l} &=& 
  \sum_{J} \frac{1}{8} J^2 (\chi _{J} - \chi _{J-1}) \\
 &\rightarrow&
  \sum_{J} \frac{1}{8} J^2
  (\chi _{J} e^{-J(J+1)t} - \chi _{J-1} e^{-(J-1)Jt}),
\end{eqnarray*}
which leads to the regularized result 
\begin{eqnarray}
 Z_{l} &=& 
  - \frac{1}{8} \sum_{J} (2J+1)\chi _J e^{-J(J+1)t}.
\end{eqnarray}
It is interesting to note that this link partition function coincides 
with the heat-kernel lattice gauge theory action but the interpretations 
and the origin of the terms are quite different%
\cite{Boulatov}\cite{Fukuma}\cite{Creutz}. 
This regularization factor $e^{-J(J+1)t}$ breaks
the Alexander invariance of the partition function but 
it will be recovered at the end of the calculation when 
we take the limit $t \rightarrow 0$. 

One of the remarkable points of this result is that 
the factor $2J+1$ has appeared in the link partition function $Z_l$, 
which is the same and necessary factor for the P-R model to assure the 
Alexander invariance.
Another important point is that the character has appeared after the 
$e^a$ integration, which makes $U$ integration straightforward. 

\subsection{ $U$ integration}

After $e^a$ integration and dividing the unimportant constant factor 
$\prod_{l} (-1/8)$, the partition function leads 
\begin{equation}
 Z = \int {\cal D}U \prod_{l} \sum_{J=0}^{\infty} \left(2J+1\right)
  \chi_{J}\Bigl(\plaq \Bigr) ~ e^{-J(J+1)t}, 
\end{equation}
where we take $t\rightarrow 0$ limit in the end of calculation.
We now carry out $DU$ integration of this partition function. 
Thanks to the character of the partition function, $DU$ integration is 
straightforward. 
We show that the Ponzano-Regge partition function will be reproduced after 
$DU$ integration with 6-$j$ symbols together with correct coefficients and 
sign factors. 

Before getting into the details we figure out how 6-$j$ symbols appear. 
The character in the partition function is a product of $D$-function
around the boundary of dual plaquette associated to a original link. 
Each tetrahedron has six original links and there are two dual links which 
is a part of a product on the boundary of the dual plaquette associated 
to each original link.
In other words three dual links associated to a $DU$ integration thrust into 
each triangle from the center of the tetrahedron. 
Therefore twelve dual links are associated to a tetrahedron.
Each $DU$ integration of the product of three $D$-function reproduces two 
3-$j$ symbols, thus we get eight 3-$j$ symbols for each tetrahedron. 
Four out of eight 3-$j$ symbols lead to a 6-$j$ symbol and the rest of four 
3-$j$ symbols lead to give trivial factor together with the 3-$j$ symbols 
from the neighbouring tetrahedra.

We first note that the character appearing in the partition function is 
a product of $D$-functions
\begin{equation}
 \chi_J(|F|) = \chi_J \Bigl(\plaq \Bigr)
             = D^J_{m_1m_2}(U_1)D^J_{m_2m_3}(U_2)\cdots 
               D^J_{m_km_1}(U_k),
\end{equation}
where $U_i$ is a dual link variables on the boundary of dual plaquette 
$\tilde P(l)$ associated to a link $l$ and $m_i$ is the third component of 
angular momentum $J$ which is assigned to the link $l$. 
As we have already pointed out that the direction of $U_i$ for each link 
is defined inward for each tetrahedron. 
On the other hand the direction of the loop composed of the product of 
dual links associated to the link $l$ can be chosen arbitrarily. 
Therefore some of $U_i$ in the above $D$-functions are $U_i^\dagger$. 
If the original link $l$ is a link of a particular tetrahedron, 
two $D$-functions out of the above product are located inside the 
tetrahedron.

\begin{figure}
\begin{center}
 \begin{minipage}[b]{0.4\textwidth}
 \epsfxsize=\textwidth \epsfbox{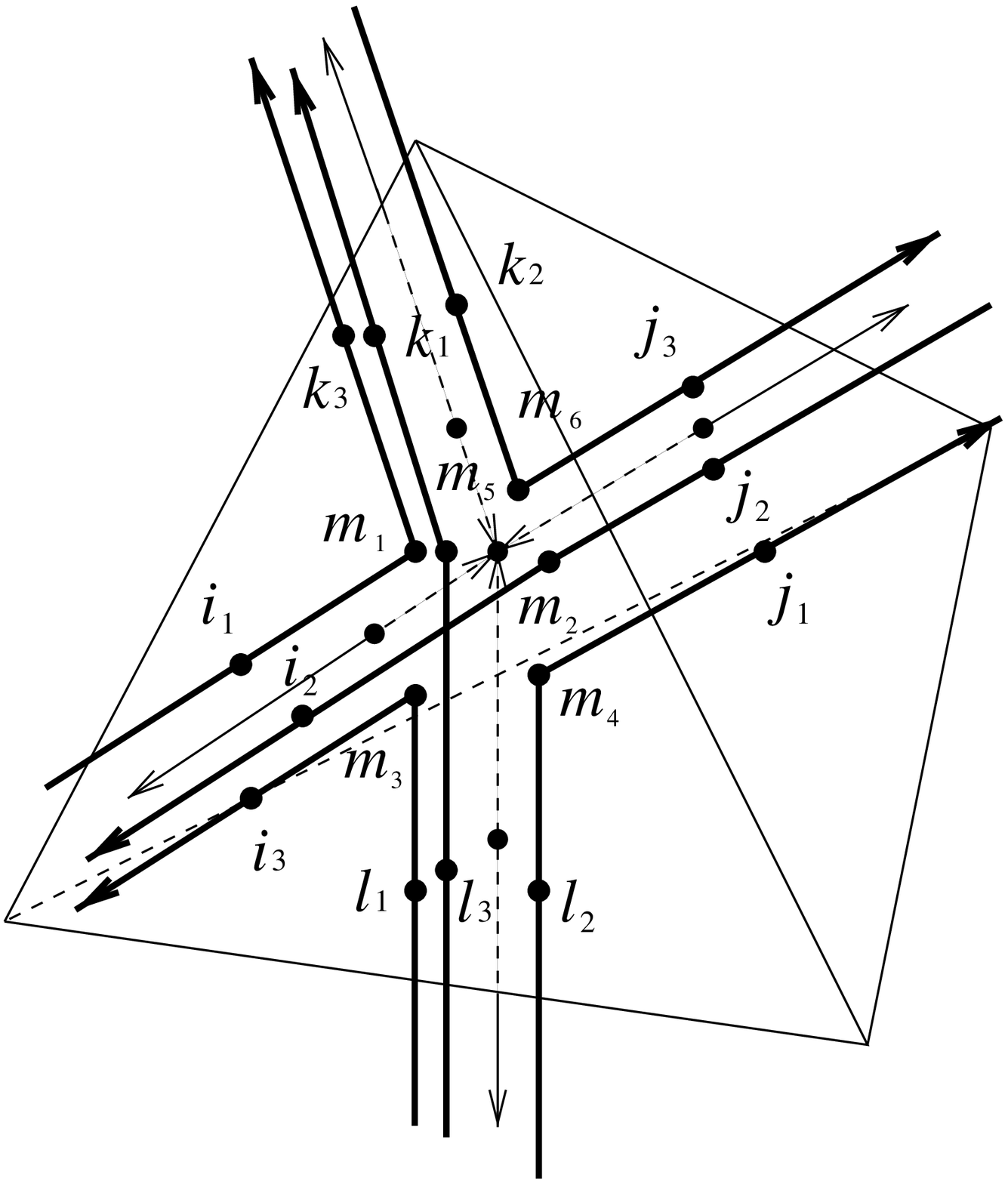} 
 \end{minipage}
 \hspace*{5mm}
 \begin{minipage}[b]{0.4\textwidth}
 \epsfxsize=\textwidth \epsfbox{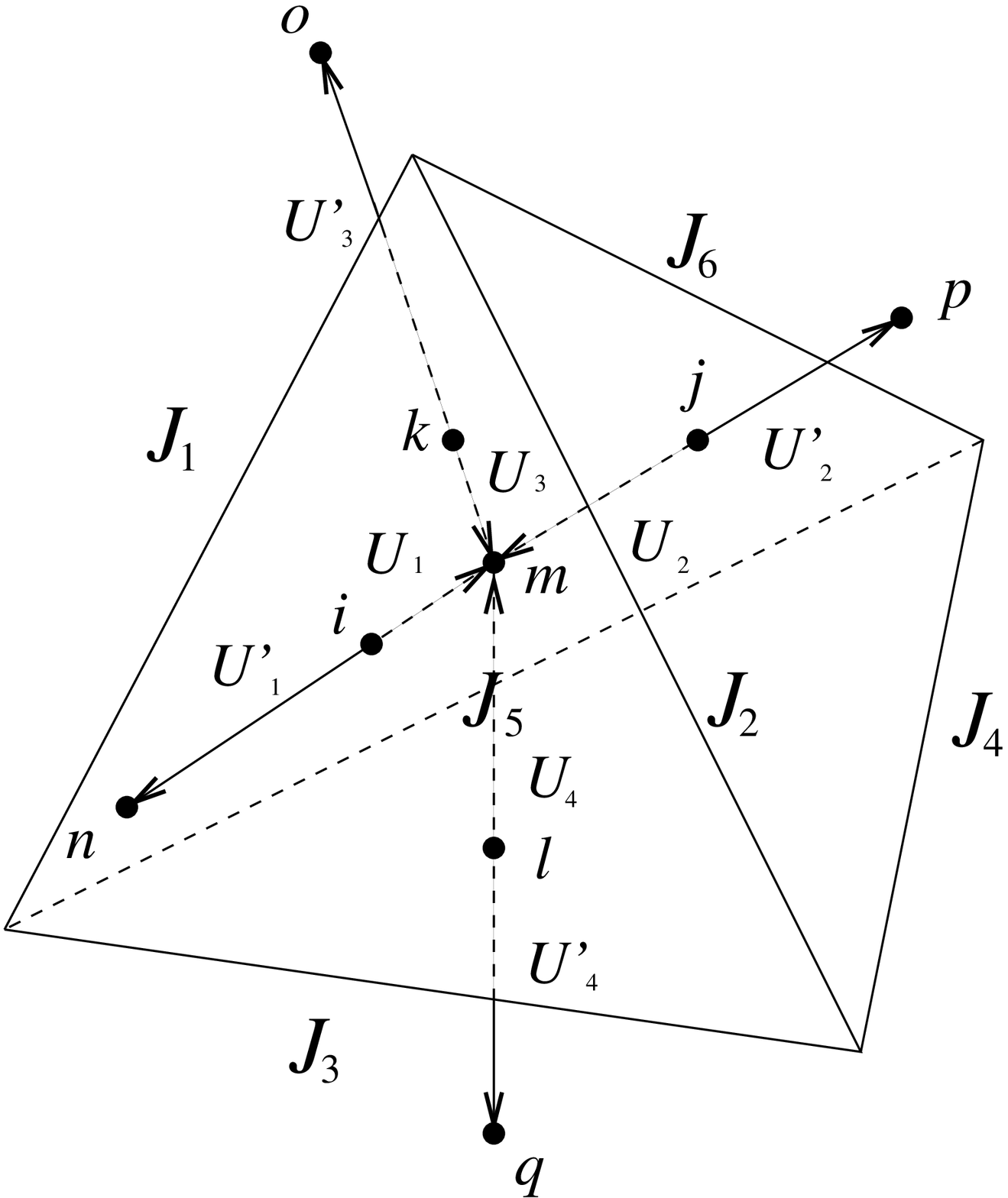} 
 \end{minipage}
\end{center}
 \caption{dual links related to neighbouring tetrahedra and the orientability}
 \label{fig:U-int}
\end{figure}

We now choose a particular situation which is shown in Fig.\ref{fig:U-int}. 
The twelve $D$-functions associated to this particular tetrahedron are  
\begin{eqnarray*}
&& I_{U_1U_2U_3U_4} = \int \prod_{i=1}^{4} DU_i ~
  D^{J_1}_{i_1m_1}(U_1) D^{J_1}_{m_1k_3}(U_3^{\dagger}) \cdot
  D^{J_2}_{j_2m_2}(U_2) D^{J_2}_{m_2i_2}(U_1^{\dagger}) \\
&& \quad\quad\quad\quad \times
  D^{J_3}_{l_1m_3}(U_4) D^{J_3}_{m_3i_3}(U_1^{\dagger}) \cdot
  D^{J_4}_{l_2m_4}(U_4) D^{J_4}_{m_4j_1}(U_2^{\dagger}) \\
&& \quad\quad\quad\quad \times
  D^{J_5}_{l_3m_5}(U_4) D^{J_5}_{m_5k_1}(U_3^{\dagger}) \cdot
  D^{J_6}_{k_2m_6}(U_3) D^{J_6}_{m_6j_3}(U_2^{\dagger}).
\end{eqnarray*}
We pick up the $D$-functions associated to $DU_1$ integration 
\begin{equation}
 I_{U_1}= (-)^{i_2-m_2+i_3-m_3}
 \int DU_1 D^{J_1}_{i_1m_1}(U_1)D^{J_2}_{-i_2-m_2}(U_1)
               D^{J_3}_{-i_3-m_3}(U_1), 
\end{equation}
where we have used the following formula to rewrite only with $U_1$ variable:
\begin{equation}
 D^I_{mn} (U^{\dagger}) =
 D^{I *}_{nm} (U) = (-)^{n-m} D^I_{-n-m} (U) 
  \label{D conjugate}.
\end{equation}
We can now use the formula relating the integration of
three $D$-functions and two 3-$j$ symbols\cite{Angular},
\begin{equation}
 \int DU D^{I}_{m_1n_1}(U) D^{J}_{m_2n_2}(U) D^{K}_{m_3n_3}(U) \nonumber \\ 
= \threej{I}{J}{K}{m_1}{m_2}{m_3}\threej{I}{J}{K}{n_1}{n_2}{n_3},
\label{3D int}
\end{equation}
which leads to the result of $DU_1$ related integration 
\begin{equation}
  I_{U_1}= (-)^{i_2-m_2+i_3-m_3} \threej{J_1}{J_2}{J_3}{m_1}{-m_2}{-m_3}
                        \threej{J_1}{J_2}{J_3}{i_1}{-i_2}{-i_3}. 
\end{equation}
After carrying out $DU_2DU_3DU_4$ integration, we obtain
\begin{eqnarray*}
I_{U_1U_2U_3U_4}
&=&
  (-)^{i_2-m_2+i_3-m_3} \threej{J_1}{J_2}{J_3}{m_1}{-m_2}{-m_3}
                        \threej{J_1}{J_2}{J_3}{i_1}{-i_2}{-i_3} \\
&\times&  (-)^{j_1-m_4+j_3-m_6} \threej{J_4}{J_2}{J_6}{-m_4}{m_2}{-m_6}
                        \threej{J_4}{J_2}{J_6}{-j_1}{j_2}{-j_3} \\
&\times&  (-)^{k_3-m_1+k_1-m_5} \threej{J_1}{J_5}{J_6}{-m_1}{-m_5}{m_6}
                        \threej{J_1}{J_5}{J_6}{-k_3}{-k_1}{k_2} \\
&\times&  \hspace{3cm}        \threej{J_4}{J_5}{J_3}{m_4}{m_5}{m_3}
                        \threej{J_4}{J_5}{J_3}{l_2}{l_3}{l_1}.
\end{eqnarray*}
We now use the formula,
\begin{eqnarray*}
 \sixj{J_1}{J_2}{J_3}{J_4}{J_5}{J_6} 
&=& \sum_{\hbox{\tiny{all $m_i$}}}
 (-1)^{\sum_i (J_i - m_i)}
 \threej{J_1}{J_2}{J_3}{-m_1}{-m_2}{-m_3}\nonumber \\
&& \hspace{-1cm} \times 
 \threej{J_1}{J_5}{J_6}{m_1}{-m_5}{m_6} 
 \threej{J_4}{J_2}{J_6}{m_4}{m_2}{-m_6}
 \threej{J_4}{J_5}{J_3}{-m_4}{m_5}{m_3},
\end{eqnarray*}
for the four 3-$j$ symbols which carry $m_i$ suffices associated to 
the center of the tetrahedron.
 We then find 6-$j$ symbols after $DU_1DU_2DU_3DU_4$ integration
\begin{eqnarray}
I_{U_1U_2U_3U_4}
&=&
  (-)^{\sum_{i=1}^{6} J_i} \sixj{J_1}{J_2}{J_3}{J_4}{J_5}{J_6} \nonumber \\
&\times&
  (-)^{i_2+i_3} \threej{J_1}{J_2}{J_3}{i_1}{-i_2}{-i_3} 
  (-)^{j_3+j_1} \threej{J_4}{J_2}{J_6}{-j_1}{j_2}{-j_3} \nonumber \\
&\times&
  (-)^{k_3+k_1} \threej{J_1}{J_5}{J_6}{-k_3}{-k_1}{k_2}
                 \threej{J_4}{J_5}{J_3}{l_2}{l_3}{l_1}.
                 \label{6jwith3j}
\end{eqnarray}

Here we are considering $SO(3)$ case then the third component of the 
angular momentum $m_i$ is integer and thus we can use the relation 
$(-)^{m_i}=(-)^{-m_i}$ which is not correct if $m_i$ is half integer in 
case of $SU(2)$.
We now look at the rest of the 3-$j$ symbols in 
eq.(\ref{6jwith3j}) which carry the 
suffices $i,j,k,l$. 
As we can see from Fig.\ref{fig:U-int}
that $DU_1$ integration reproduces two 
3-$j$ symbols and one of them associated to the suffices $m_k$ is 
absorbed to reproduce the 6-$j$ symbol and the other 3-$j$ carrying the 
suffices $i_k$ could be combined with another 3-$j$ symbol obtained from 
$DU'_1$ integrations of the neighbouring tetrahedron. 
Those 3-$j$ symbols are associated to the boundary triangle of the 
two neighbouring tetrahedron carrying suffix $i_k$. 
In this particular case of Fig.\ref{fig:U-int} we obtain the following two 3-$j$ symbols
\begin{eqnarray*}
I^b_{i}
=
  \sum_{i_1i_2i_3}(-)^{i_1+i_2+i_3} 
   \threej{J_1}{J_2}{J_3}{i_1}{-i_2}{-i_3} 
   \threej{J_1}{J_2}{J_3}{i_1}{-i_2}{-i_3}. 
\end{eqnarray*}
Since the three angular momentum vectors $J_1,J_2,J_3$ construct the 
boundary triangle, the third components satisfy the relation 
$i_1-i_2-i_3=0$.
Using the following formula:
\begin{equation}
 \sum_{m_1m_2m_3}
 \threej{J_1}{J_2}{J_3}{m_1}{m_2}{m_3}
 \threej{J_1}{J_2}{J_3}{m_1}{m_2}{m_3} = 1, 
\end{equation}
and noting $(-)^{i_1+i_2+i_3}=(-)^{i_1-i_2-i_3}=1$ for $SO(3)$ case, 
these two 3-$j$ symbols lead to a trivial factor. 

It should be pointed out here that the above factor reproduces a negative sign 
if $i_k$ is half integer in case of $SU(2)$ since 
$(-)^{i_1+i_2+i_3}=(-)^{2i_1}(-)^{-i_1+i_2+i_3}=-1$.
Some of negative sign factors can be removed by using the triangle relation 
and $(-)^{m_1-i_1}=(-)^{-m_1+i_1}$, which holds even for half integer 
values since $m_1$ and $i_1$ are both the 
third components of $J_1$. 
We cannot, however, get rid of all the negative sign factors
for the dual link loop associated to the original link in this way. 
Therefore we need to point out that there appear floating 
negative sign factors for $SU(2)$ case.  

Finally we have found that our partition function is the same as
that of the Ponzano-Regge model, except for the regularization factor 
$\Lambda(\lambda) = \sum_{J=0}^{\lambda}(2J+1)^2$.
Since our partition function is divergent with the same reason, 
we should introduce the same regularization factor as
the P-R model.

\setcounter{equation}{0}

\section{Conclusion and the Interpretations}

We have proposed the partition function of the $ISO(3)$ 
lattice Chern-Simons action  
\begin{eqnarray*}
 Z_{\hbox{\scriptsize LCS}}
  &=& \lim_{t \rightarrow 0} \lim_{\lambda \rightarrow \infty}
  \int \frac{{\cal D}U {\cal D}e}
  {\displaystyle \prod_{l} \Bigl(-\frac{1}{8} \Bigr)
  \prod_{\hbox{\tiny verticies}}\Lambda(\lambda) }
 \int {\cal D}U' K(U,U';t)   \\
&&  \times \frac{e^3}{|e|} \left[ \prod_{a=1}^{2}
   \delta \left( \frac{F^a}{F} + \frac{e^a}{e} \right)
  + \prod_{a=1}^{2}
  \delta \left( \frac{F^a}{F} - \frac{e^a}{e} \right)
  \right]
  \sum_{J=0}^{\lambda} \delta \left( e - \frac{J}{2} \right)
  e^{i S_{\hbox{\tiny LCS}} (e, U')}
\end{eqnarray*}
which exactly coincides with the Ponzano-Regge model 
after the integration of the dreibein and the dual link variables.
The discreteness of the length of the dreibein is the natural consequence 
of the logarithm form in the lattice Chern-Simons action.
On the simplicial lattice manifold constructed from tetrahedra, 
the dreibeins are located on the original links while the lattice version of 
the spin connection, the dual link variables are located on the dual links.
We have explicitly shown that the lattice Chern-Simons action is invariant 
under the lattice version of local Lorentz transformation and the 
lattice gauge diffeomorphism.
In order to get the topological gravity theory at the quantum level, 
we need a constraint which solves the spin connection as a function of 
the dreibein.
We have found the constraint which can be interpreted as the gauge fixing 
condition of the lattice gauge diffeomorphism. 

Since the Ponzano-Regge model is invariant under the 2-3 and 1-4 Alexander 
moves, the partition function is invariant under how the three dimensional 
space is divided into small pieces by tetrahedra.
It is natural to expect that the partition function is invariant in the 
continuum limit and the lattice Chern-Simons action leads to the 
continuum Chern-Simons action. 

It is interesting to note that the algebraic dual nature of the one forms 
$e$ and $\omega$ in the $BF$ theory or equivalently in the Palatini action, 
is reflected on the geometric dual nature of $e$ and 
$U=\hbox{e}^\omega$ on the lattice.
In other words, $e$ and $\omega$ are defined
in the Lie algebra ${\cal L}^*_G$ and 
${\cal L}_G$, respectively, where ${\cal L}^*_G$ and ${\cal L}_G$ are 
dual to each other with $G=SO(3)$ and 
${\cal L}^*_G \oplus {\cal L}_G=ISO(3)$\cite{Ashtekar}. 
The one form $e$ and the link variable $U=\hbox{e}^\omega$ are defined on 
a original links and dual links, respectively, and thus the dual nature 
of the algebra is reflected in the geometry on the lattice.  

In the $ISO(3)$ lattice Chern-Simons action there are 6 gauge parameters. 
Two gauge parameters of the lattice gauge diffeomorphism can be used to 
rotate the dreibein $e^a$ to be parallel or anti-parallel to the curvature 
$F^a$ and the remaining one gauge parameter of the lattice gauge 
diffeomorphism can be exhausted to make the length of the dreibein discrete. 
There remain three gauge parameters of the lattice local Lorentz gauge 
symmetry, which are expected to convert into the three vector parameters 
of general coordinate diffeomorphism symmetry. 
There are two reasons to expect this scenario. 
Firstly the lattice action coincides with the Ponzano-Regge model 
which is Alexander move invariant and is thus expected to be metric 
independent. 
In fact the lattice Chern-Simons action in the continuum limit is metric 
independent since it is composed of one form.  
Secondly the general coordinate transformation of diffeomorphism and the 
local Lorentz transformation are on shell equivalent in the continuum 
$ISO(3)$ Chern-Simons gravity\cite{Witten}.  

In this paper we have concentrated on the relation between the $ISO(3)$
Chern-Simons gravity and Ponzano-Regge model. 
The $q$-deformed Ponzano-Regge model proposed by Turaev and 
Viro\cite{Turaev-Viro} is expected to be related with $SO(4)$ or $SO(3,1)$ 
Chern-Simons gauge theory and lead to Einstein gravity 
with a cosmological term\cite{Witten}\cite{Ooguri-Sasakura}. 
It is thus natural to extend the present formulation into the 
lattice gauge gravity with cosmological term and try to find the 
connection with the $q$-deformed version of Ponzano-Regge model.
There are already several trials on these directions%
\cite{Tada}\cite{Boulatov2}
but our lattice formulation may give new insights.   

Since our lattice gauge gravity formulation of the present paper 
has natural correspondence with the Regge calculus\cite{Kawamoto-Nielsen},  
it is expected that the extension to four dimension is 
straightforward and is expected to be related with the $BF$ gravity theory 
which has 15-$j$ interpretation\cite{Ooguri}.
As in the present formulation the area of the triangle will be discretized 
in four dimensional lattice gauge gravity theory\cite{Smolin}.

It will be interesting to extend the current formulation into the lattice 
version of generalized Chern-Simons gravity. 
The three dimensional generalized Chern-Simons action includes not only 
one form gauge field but also zero, two and three form gauge fields%
\cite{Kawamoto-Watabiki1}.
It will be interesting to ask what the role of other form gauge fields is. 
It should also be noted that the four dimensional generalized Chern-Simons 
action includes $BF$ term together with several other terms which 
include zero, one, two, three, and four forms.

\vskip 1cm

\noindent{\Large{\bf Acknowledgements}} \\
We thank M.Fukuma, T.Kugo, L.Smolin and H.Suzuki for useful discussions 
and comments. 
This work is supported in part by Japan Society for the Promotion of 
Science under the grant number 09640330.

 
\end{document}